\documentstyle[11pt]{article}

\begin{document}

\noindent
{\bf Comment on ``Entropy Generation in Computation and the Second law of
Thermodynamics'', by S. Ishioka and N. Fuchikami }
({\tt chao-dyn/9902012} 17 Feb 1999).
\medskip

In the above cited paper, the authors claim that a more precise expression
of Landauer's principle, namely:\ ``[L] erasure of information is
accompagnied by {\em heat }generation to the amount of $kT\ln 2$/bit;''
should be:\ ``[IF] erasure of information is accompagnied by {\em entropy}
generation $k\ln 2$/bit.''\ However, as they probably ignore, Landauer's
statement about heat dissipation in computation has been {\it a priori}
derived from phase space contraction arguments which can be stated
equivalently in terms of entropy. Hence, [L] and [IF] are equivalent, even
from Landauer's viewpoint, and there's no need to argue for a difference.

To clarify this point, let us consider the example of the bistable potential
(cf.\ cited paper). According to Ishioka and Fuchikami, entropy is generated
when a particle, initially trapped in one side of the bistable potential, is
brought to a state where it can move freely between the two wells by
lowering the potential barrier. Logically, this leads to the erasure of the
information contained in the ``position'' bit (0 or 1 depending on the initial state
in the double well), and, physically, to an increase of the entropy of the
particle from $S_i=0$ (definite memory state)\ to $S_f=\ln 2$ (unknown final
position). Hence, their conclusion [IF].

In Landauer's analysis of the problem on the other hand, a similar
conclusion is reached with the difference that the erasure action is
somewhat inversed. Landauer considers a particle in a bistable potential
which is used to record the position of a particle in another bistable potential.
From an outside point of view, the outcome of one measurement of the
position generates a random bit that is stored in the recording bistable
system. Hence, in the process, the recording particle goes from a definite
state of zero entropy (the ready-for-measurement state) to a record state
which is 0 with some probability $p$ and 1 with probability $1-p$, thereby
increasing the entropy of the recording device. The erasure of the information
then proceeds by putting the recording particle back to its
ready-for-measurement state. This leads to a
dissipation of {\it entropy} conveyed as a dissipation of {\it heat} ([L]).

The two exposed viewpoints differ only by the role
of the recording device and the position taken in the act of erasure. In
Zurek's terminology ({\it Phys. Rev. }A 40(8) 4731, 1989), Ishioka and Fuchikami
analyze their system from an inside point of view which assigns no
probabilities for the recording state, whereas the second analysis considers
the measured system and the recording system from the same probabilistic
perspective. Evidently, both are correct and lead to the same result,
namely, that {\it entropy is generated in the erasure process}. The dissipation of
heat is just an {\em a posteriori} conclusion which follows from the
connection with thermodynamics.

To conclude, let us also note that the ``writing process'' discussed in the
article of Ishioka and Fuchikami can be performed without doing any work. 
In Szilard's engine, for example, one can insert the partition after waiting until 
the particle goes on the side that corresponds to the bit to be registered. 

\bigskip

Hugo Touchette

MIT, {\tt htouchet@mit.edu}

02/19/1999

\end{document}